\documentclass[entropy,article,accept,moreauthors]{Definitions/mdpi}

\firstpage{1}
\pubvolume{0}
\issuenum{0}
\articlenumber{65}
\pubyear{2022}
\copyrightyear{2022}
\history{Received: date; Accepted: date; Published: date}

\usepackage{color,xspace}
\usepackage{Definitions/macros_gpi}
\usepackage{aas_macros}
\usepackage{amssymb}
\usepackage[nameinlink,capitalize,noabbrev]{cleveref}
\usepackage{comment}

\usepackage{subcaption}

\graphicspath{{Figs/}}

\setitemize{parsep=6pt,itemsep=0pt,leftmargin=*,labelsep=5.5mm}
\setenumerate{parsep=6pt,itemsep=0pt,leftmargin=*,labelsep=5.5mm}
\setlist[description]{itemsep=0mm}
\usepackage{environ}
\NewEnviron{myequation}{%
\begin{equation}
  \scalebox{0.95}{$\BODY$}
\end{equation}
}

\Title{\texttt{SuperNest}: accelerated nested sampling applied to astrophysics and cosmology$^\dagger$}

\Author%
{%
  Aleksandr Petrosyan$^{1,2,3}$\thanks{E-mail: ap886@cam.ac.uk}\:\: \& %
  Will Handley$^{1,2,4}$\thanks{E-mail: wh260@cam.ac.uk}%
}
\AuthorNames{Aleksandr Petrosyan and Will Handley}

\address{%
  $^{1}$Astrophysics Group, Cavendish Laboratory, J. J. Thomson Avenue, Cambridge, CB3 0HE, UK\\
  $^{2}$Kavli Institute for Cosmology, Madingley Road, Cambridge CB3 0HA, UK\\
  $^{3}$Queens' College, Silver Street, Cambridge, CB3 9ET, UK\\
  $^{4}$Gonville \& Caius College, Trinity Street, Cambridge, CB2 1TA,UK
}

\firstnote{Submitted to International Workshop on Bayesian Inference and Maximum Entropy Methods in Science and Engineering, IHP, Paris, July 18--22, 2022.}

\abstract{ We present a method for improving the performance of nested
  sampling as well as its accuracy. Building on previous work by
  \citet{chen-ferroz-hobson}, we show that posterior repartitioning
  may be used to reduce the amount of time nested sampling spends in
  compressing from prior to posterior if a suitable ``proposal''
  distribution is supplied. We showcase this on a cosmological example
  with a Gaussian posterior, and release the code as an LGPL licensed,
  extensible Python package
  \href{https://gitlab.com/a-p-petrosyan/sspr}{\texttt{supernest}}.  }

\begin{document}


\section{Introduction}\label{sec:introduction}

Nested sampling is a multi-purpose tool invented by John
\citet{Skilling2006} for simultaneous probabilistic integration of
high-dimensional distributions and drawing weighted samples. These are
essential tasks for the performance of Bayesian model comparison and
parameter estimation~\cite{importanceOfZ}. It has become widely used
across the physical sciences from astrophysics \& cosmology through
biochemistry and materials science to high energy physics~\cite{NRP}.

The performance of nested sampling algorithms however suffer in
comparison with alternative posterior samplers such as
Metropolis-Hastings~\citep{Metropolis}, in particular when the latter
are provided with a pre-trained proposal distribution and start
point. This paper allows an apples-to-apples comparison, by
providing a mechanism for accelerating nested sampling using proposal
distributions. This is achieved using an extension of the posterior
repartitioning techniques developed by~\citet{chen-ferroz-hobson}.

In \cref{sec:background} we establish notation and definitions. In
\cref{sec:ppr} we review and extend posterior repartitioning. In
\cref{sec:supernest} we present our chosen approach and in
\cref{sec:results} we showcase this on a cosmology example before
concluding in \cref{sec:conclusions}.  \pagebreak

\section{Background}\label{sec:background}
\subsection{Notation}
In this section we establish notation and recall the key terms in
nested sampling and Bayesian inference.  A probabilistic inference
model \(\mathcal{M}\) with a vector of tunable parameters \(\theta\)
for data \(D\) is specified by \emph{likelihood}
\(\mathcal{L}(\theta)\equiv P(D|\theta,\mathcal{M})\) and \emph{prior}
\(\pi(\theta) = P(\theta|\mathcal{M})\). The two-step challenge of
inference is to compute the \emph{posterior}
\(\mathcal{P}(\theta)\equiv P(\theta|D,\mathcal{M})\) and
\emph{evidence} \(\mathcal{Z}=P(D|\mathcal{M})\). Constructing a
representation of the posterior is termed \emph{parameter estimation},
and evaluating an estimate of the evidence comprises \emph{model
  comparison}. The four distributions are linked via Bayes theorem:
\begin{equation}\label{eq:bayes}
    {\cal L}(\theta) \times \pi (\theta) = {\cal Z} \times {\cal P} (\theta),
\end{equation}
where both sides of the equation equal to the joint distribution \(P(D,\theta)\).

It is useful to define the \emph{Kullback-Leibler divergence}
\(\mathcal{D}\) \citep{Kullback_1951} between the prior \(\pi\) and
the posterior \(\mathcal{P}\):
\begin{equation}\label{eq:kl-def}
  \mathcal{D}_{\pi}\{\mathcal{P} \} = \int \mathcal{P}(\theta) \ln \frac{\mathcal{P}(\theta)}{\pi(\theta)} d \theta.
\end{equation}
One can understand the properties of the KL divergence as a
logarithmic ratio of the volume contained in the prior distribution to
the volume contained in the posterior distribution:
\begin{equation}\label{eq:4}
  {\mathcal{D}_{\pi}\{ \mathcal{P} \} \approx \ln \frac{V_{\pi}}{V_{\mathcal{P}}}}.
\end{equation}
When both the prior and posterior are uniform (i.e.~constant value
where they are non-zero), this relation is exact\footnote{One can think of the
  posterior averaging process in \cref{eq:kl-def} as an information
  theory inspired ``smoothing'' of this process for non top-hat
  distributions.}.

\subsection{Nested Sampling}\label{sec:ns-primer}

In this section we'll provide a brief primer on the nested sampling
algorithm. For a survey of established nested sampling packages and
applications, readers are recommended the Nature primer~\cite{NRP}.
We begin by initialising \(n_\mathrm{live}\) initial ``live points''
drawn from the prior \(\pi\).  Then at each iteration the lowest
likelihood live point is marked ``dead'' and replaced with a new live
point drawn from the prior, that is at higher likelihood.  The
compression procedure allows one to estimate the density of states,
which given the recorded locations of live and dead points leads to
estimates of posterior and evidence with respective errors.

Crucially, the algorithm terminates once the running estimate of the
error in the evidence is greater than a set fraction of the total
estimated evidence. Thus, the time complexity of nested sampling is:
\begin{equation}\label{eq:3}
  T \propto n_\text{live}\times  \langle \mathcal{T}\{\mathcal{L}(\theta)\} \rangle\times \langle \mathcal{T}\{\text{Impl.}\} \rangle \times \mathcal{D}_{\pi}\{\mathcal{P}\},
\end{equation}
where \(\mathcal{T}\{\mathcal{L}(\theta)\}\) is the time complexity of
evaluating \(\mathcal{L}(\theta)\) and \(\mathcal{T}\{\text{Impl.}\)
is the implementation-specific factor representing the time complexity
of replacing a dead point with a live point with a higher likelihood,
and is the dominant factor for nested sampling over parameter spaces
with many dimensions \citep{NRP}. For rejection nested sampling
(e.g. \texttt{MultiNest}~\cite{Feroz2009MultiNestAE}) this factor is
exponential in the number of tunable parameters \(d\) (with a turnover
constant in tens of dimensions), and linear in \(d\) for chain-based
approaches (e.g. \texttt{PolyChord}~\cite{polychord}).  The
Kullback-Leibler divergence \(\mathcal{D}_\pi\{\mathcal{P}\}\) is
linear in the \(d\) for parameters that are well constrained, as can
be seen from \cref{eq:4} by noticing that volumes in parameter space
scale with a power of \(d\).

From \cref{eq:3} it follows that the run time of nested sampling,
given a resolution fixed by \(n_\mathrm{live}\) can be reduced by (i)
speeding up the likelihood evaluation; (ii) improving the replacement
efficiency; (iii) speeding up the prior to posterior compression. The
final approach is the one adopted for the \emph{repartitioning
  techniques} in this paper.

The accuracy of the evidence is associated with the accumulated
Poisson noise, such that the error in \(\ln \mathcal{Z}\)
is~\cite{Skilling2006}:
\begin{equation}
  \sigma \propto \sqrt{\mathcal{D}_\pi\{\mathcal{P}\}/n_\mathrm{live}}.
\end{equation}
Thus, faster compression leads to better accuracy and precision. For a
fixed accuracy \(\sigma\), we can adjust
\({n_\mathrm{live}\propto \mathcal{D}/\sigma^2}\), which means
\({T \propto \mathcal{D}^2/\sigma^2}\), given \cref{eq:3}. In precision-normalised cases,
reducing the effect of the KL divergence term results in quadratic
improvements to the speed of nested sampling.  Note that the number of
live points can be also adjusted. The minimum number is dictated by
the quantity of modes in the multi-modal posterior, and implementation dependent considerations: for example, \texttt{PolyChord}
needs \({n_\mathrm{live}\sim\mathcal{O}(d)}\) to generate enough dead point phantoms
for an accurate covariance computation.  \texttt{MultiNest} needs at least
\({n_\mathrm{live} \sim \mathcal{O}(100)}\) to compute ellipsoidal decompositions.

\subsection{Historical overview of acceleration attempts}\label{sec:note-other-optim}

One approach to partially achieve this form of speed-up is to perform
two passes of nested sampling. First a low resolution pass (few live
points) over the entire original prior volume \(V_{\pi, 1}\) finds the
location of the highest concentration of posterior volume, and an
evidence estimate \(\mathcal{Z}_{1}\).  Next, one constructs a much
tighter bounding prior \(\pi_2\) with volume \(V_{\pi, 2}\) around the
largest concentration of the posterior.  Trivially:
\begin{equation}\label{eq:5}
  \mathcal{D}_{\pi, 1} \{ \mathcal{P} \} < \mathcal{D}_{\pi, 2} \{ \mathcal{P} \}
\end{equation}
hence latter inference over the tighter box will finish faster finding
the evidence \(\mathcal{Z}_{2}\). With a changed prior, this evidence will not
be the same as the original. However, for box-priors we can compensate
the volumes geometrically, using the fact that the overall evidence
satisfies:
\begin{equation}\label{eq:6}
  \mathcal{Z} = \mathcal{Z}_{1} = \mathcal{Z}_{2} \frac{V_{\pi, 2}} {V_{\pi, 1}}
\end{equation}
See~\citet{hacky} for examples of this method's application in real
astronomical cases.

\section{Posterior repartitioning}\label{sec:ppr}

Nested sampling is unusual amongst Bayesian numerical algorithms, in
that it distinguishes between the prior \({\pi(\theta)}\) and likelihood
\({\mathcal{L}(\theta)}\). Metropolis Hastings, and Hamiltonian
Monte-Carlo for example are techniques which are only sensitive
\({\mathcal{L} \times \pi}\) (i.e.~the joint or unnormalised posterior) is
considered for the purposes of acceptance/rejection. Nested sampling
on the other hand by ``sampling from the prior $\pi$, subject to a
hard likelihood constraint \(\mathcal{L}\)'' separates the two
functions.

\citet{chen-ferroz-hobson} pioneered an approach which exploits this
split, noting that we have the freedom to define a new prior
\(\tilde{\pi}\) and likelihood \(\tilde{\mathcal{L}}\), which providing
they obey the relation:
\begin{equation}
    \tilde{\mathcal{L}}(\theta) \tilde{\pi}(\theta) = \mathcal{L}(\theta) \pi(\theta)
\end{equation}
will recover the same posterior and evidence. Conceptually one can think of
this as transferring portions of the likelihood into the prior, or
vice versa. A cosmological example of this subtle rearranging of
portions of likelihood into prior can be found in many codes which
implement nested sampling such as
\texttt{cobaya}~\citep{2021JCAP...05..057T},
\texttt{cosmomc}~\citep{2002PhRvD..66j3511L,2013PhRvD..87j3529L},
\texttt{MontePython}~\citep{2019PDU....24..260B} or
\texttt{cosmosis}~\citep{2015A&C....12...45Z} in the manner in which
they consider Gaussian priors. Codes can either make the choice to
incorporate this as part of the prior (typically as an inverse error
function based unit hypercube transformation), or as a density term
added to the rest of the log-likelihood calculations.

This rearranging affects the Kullback-Leibler divergence. The
difference can be found using a posterior average:
\begin{equation}\label{eq:dkl_trans}
  \mathcal{D}_{\tilde{\pi}}\{\mathcal{P}\} = \mathcal{D}_\pi\{\mathcal{P}\} + \left\langle  \ln \frac{\tilde{\pi}}{\pi}\right\rangle_\mathcal{P}
\end{equation}
If the final term in \cref{eq:dkl_trans} is negative, repartitioning
provides a mechanism for reducing the time to convergence by
transferring a reasonable ``guess'': the proposal distribution, from
likelihood to prior, reducing the Kullback-Leibler divergence between
the prior \(\tilde{\pi}\) and the posterior
\({\tilde{\mathcal{P}} = \mathcal{P}}\).  As we've seen before, this both improves
performance and increases precision.

\subsection{Example 1: Power posterior repartitioning}

The scheme considered in \citet{chen-ferroz-hobson} is \emph{power
  posterior repartitioning} (PPR):
\begin{equation}\label{eq:ppr}
  \tilde{\pi}(\theta|\beta) = \frac{{\pi(\theta)}^\beta}{z_\beta}, \qquad
  \tilde{\mathcal{L}}(\theta|\beta) = \mathcal{L}(\theta){{\pi(\theta)}^{1-\beta}}{z_\beta}, \qquad
  z_\beta = \int {\pi(\theta)}^\beta d\theta.
\end{equation}
For \({\beta=0}\), any distribution \(\pi\) becomes uniform. If a
distribution \(\pi\) is not uniform, raising it to a power of
\(\beta\) serves to sharpen or spread the peaks, depending on whether
\(\beta\) is greater or smaller than \(1\), in analogy with the
thermodynamic inverse temperature of a Boltzmann
distribution~\citep{NRP}. For any given value of \(\beta\), a nested
sampling run with likelihood \(\tilde{\mathcal{L}}\) and prior
\(\tilde{\pi}\) recovers the same posterior samples and evidence as a run
with \(\mathcal{L}\) and prior \(\pi\).

\citet{chen-ferroz-hobson} considered this in the context of a
geophysical example. The original intention behind repartitioning was
not to improve performance, but rather as a way of making nested
sampling more resilient to over-zealously specified priors (e.g.~if a
user specifies a Gaussian prior which is narrow and/or distant from
the peak of the likelihood). Still, the performance implications are
valuable.  Substituting \cref{eq:ppr} into \cref{eq:dkl_trans} shows
that varying \(\beta\) will change the time to convergence as:
\begin{equation}
  T(\beta) \propto \mathcal{D}_{\tilde{\pi}}\{\mathcal{P}\} = \mathcal{D}_\pi\{\mathcal{P}\} + \left\langle  \ln \frac{\pi^{\beta-1}}{z_\beta}\right\rangle_\mathcal{P}
\end{equation}
Whether or not this increases or decreases the speed of nested
sampling depends on the choice of \(\beta\) and to what extent the
sharpened prior overlaps with the posterior.  \cref{eq:ppr} is valid
for any choice of \(\beta\), and while any specific choice can be useful,
\citet{chen-ferroz-hobson} demonstrated that still greater power comes
from considering it as an additional parameter to sample over,
effectively extending the parameter space to
\(\tilde{\theta}=(\theta,\beta)\) with a prior on \(\beta\). A posterior value of the
parameter \(\beta\) close to \(0\) therefore indicates that the prior was
poorly chosen, i.e.\ that there is significant \emph{tension} between
prior and posterior.

One could consider this for the purposes of acceleration by allowing
\(\beta\) to take a range wider than \([0,1]\), for example, by using
an exponential prior \({P(\beta) =
  e^{-\beta}\:[0\le\beta<\infty]}\). There is however no need to restrict
ourselves to the case of power posterior repartitioning, or indeed to
a priors \(\tilde{\pi}\), which are direct transformations of the
original \(\pi\).

\subsection{Example 0: Replacement repartitioning}

The simplest repartitioning scheme is to simply use a different prior
\(\pi_*\), but compensate for this with the adjusted likelihood:
\begin{equation}
    \tilde{\pi}(\theta) = \pi_*(\theta), \qquad \tilde{\mathcal{L}}(\theta) = \frac{\mathcal{L}(\theta)\pi(\theta)}{\pi_*(\theta)}.
\end{equation}
From \cref{eq:dkl_trans}, we can see that if \(\pi_*\) is close to
\(\mathcal{P}\) as quantified by the KL divergence, nested sampling
terminates sooner (with
\({\mathcal{D}_{\tilde{\pi}}\{\mathcal{P}\}=0}\) in the limiting case
\({\tilde{\pi}=\mathcal{P}}\)). However, a poorly specified proposal
\(\pi_*\) will pessimise the choices of new live points, thus making
the problem harder. Indeed, a poorly chosen proposal can prevent us
gaining any usable information.

\subsection{Example 2: Additive superpositional repartitioning}\label{sec:linear}
Another repartitioning scheme is to consider an additive superposition
of the original prior \(\pi\) with another normalised distribution \(\pi_*\):
\begin{equation}
    \tilde{\pi}(\theta|\beta) = (1-\beta)\pi(\theta) + \beta \pi_*(\theta), \qquad \tilde{\mathcal{L}}(\theta|\beta) = \frac{\mathcal{L}(\theta)\pi(\theta)}{(1-\beta)\pi(\theta) + \beta \pi_*(\theta)}
\end{equation}
Here \(\pi_*\) does not need to be related to the original prior, and
could for example be a Gaussian proposal distribution (often supplied
with cosmological sampling packages for common likelihood
combinations), or a tighter box prior as suggested in
\cref{sec:note-other-optim}.

This scheme can be generalised to superpositions of \(N\) alternative
distributions \(\pi_1,\ldots \pi_N\). Consider:
\begin{equation}
    \tilde{\pi}(\theta|\beta) = \left(1-\sum_{i=1}^N \beta_i\right)\pi(\theta) + \sum_{i=1}^N \beta_i \pi_i(\theta), \qquad \tilde{\mathcal{L}}(\theta|\beta) = \frac{\mathcal{L}(\theta)\pi(\theta)}{(1-\sum_{i=1}^N \beta_i)\pi(\theta) + \sum_{i=1}^N \beta_i \pi_i(\theta)}
\end{equation}
However, additive superpositional repartitioning in practice has the
issue of poor generalisation to higher dimensions and/or complex
priors due to the difficulty of implementing an analytic unit
hypercube transformation. Further work is required to fully explore
this using nested samplers that don't rely on prior point-percent
functions which produce said transformations\footnote{With thanks to
  Xy Wang for exploring the possibility of these models in the first
  instance as part of her University of Cambridge Masters
  dissertation}.

  \section{\texttt{supernest}: Stochastic superpositional repartitioning}\label{sec:supernest}

  The above examples lead us to the proposed repartitioning
  scheme. Just like with additive repartitioning we allow arbitrarily
  many, unrelated priors to be superimposed.  However, we side-step
  the multivariate point-percent function correction problem by
  sampling from each of the priors with some probability. We achieve
  this by adding an integer selection parameter \(\beta\) such that:
  \begin{equation}
    \tilde{\pi}(\theta|\beta) = \left\{
        \begin{array}{ccc}
            \pi(\theta) &:& \beta=0\\
            \pi_1(\theta) &:& \beta=1\\
            \vdots&&\vdots\\
            \pi_N(\theta) &:& \beta=N
        \end{array}
    \right.
    \qquad
    \tilde{\mathcal{L}} (\theta|\beta) = \left\{
      \begin{array}{ccc}
            \mathcal{L}(\theta) &:& \beta=0\\
            \mathcal{L}(\theta)\pi(\theta)/\pi_1(\theta) &:& \beta=1\\
            \vdots&&\vdots\\
            \mathcal{L}(\theta)\pi(\theta)/\pi_N(\theta) &:& \beta=N
        \end{array}
    \right.
\end{equation}
and then \(\beta\) is sampled over with a discrete (typically uniform)
prior \(P(\beta)\). The choice of said prior depends on the confidence
in each of the proposals.  When applied to nested sampling, at the end
of the run, one will recover a (joint) posterior on \(\beta\), which
will indicate which proposal was chosen more frequently, as the usual
Bayesian balance between the corresponding modified likelihood
\(\tilde{\mathcal{L}}\) and KL divergence
\(\mathcal{D}_{\tilde{\pi}}\{\mathcal{P}\}\) of the favoured bin.

In the additive case from \cref{sec:linear}, the primary difficulty is
in obtaining the point-percent function of the repartitioned prior.
The more priors are introduced, the harder the calculation.  For the
stochastic case, the prior of the superposition is the superposition
of the individual priors. Thus, if one knows the point percent
functions of the constituent priors, simply copying them is the
``calculation''.

Additionally, only one of the prior quantile/likelihood pairs is
evaluated for each point. Thus the time complexity of the likelihood
evaluation of a superposition of multiple likelihoods is
\(\mathcal{O}(\langle \mathcal{T}(\mathcal{L}) \rangle)\). By contrast
the equivalent time complexity for the additive superposition is the
sum of the time complexities:
\(\mathcal{O}\left(\sum \mathcal{T}(\mathcal{L})\right)\).  For
comparison, if we fed 100 identical models to the stochastic mixture,
we'd not see a performance regression, while the additive
superposition would take 100 times longer to compute.

\section{Cosmological example}\label{sec:results}

\begin{figure}
  \includegraphics{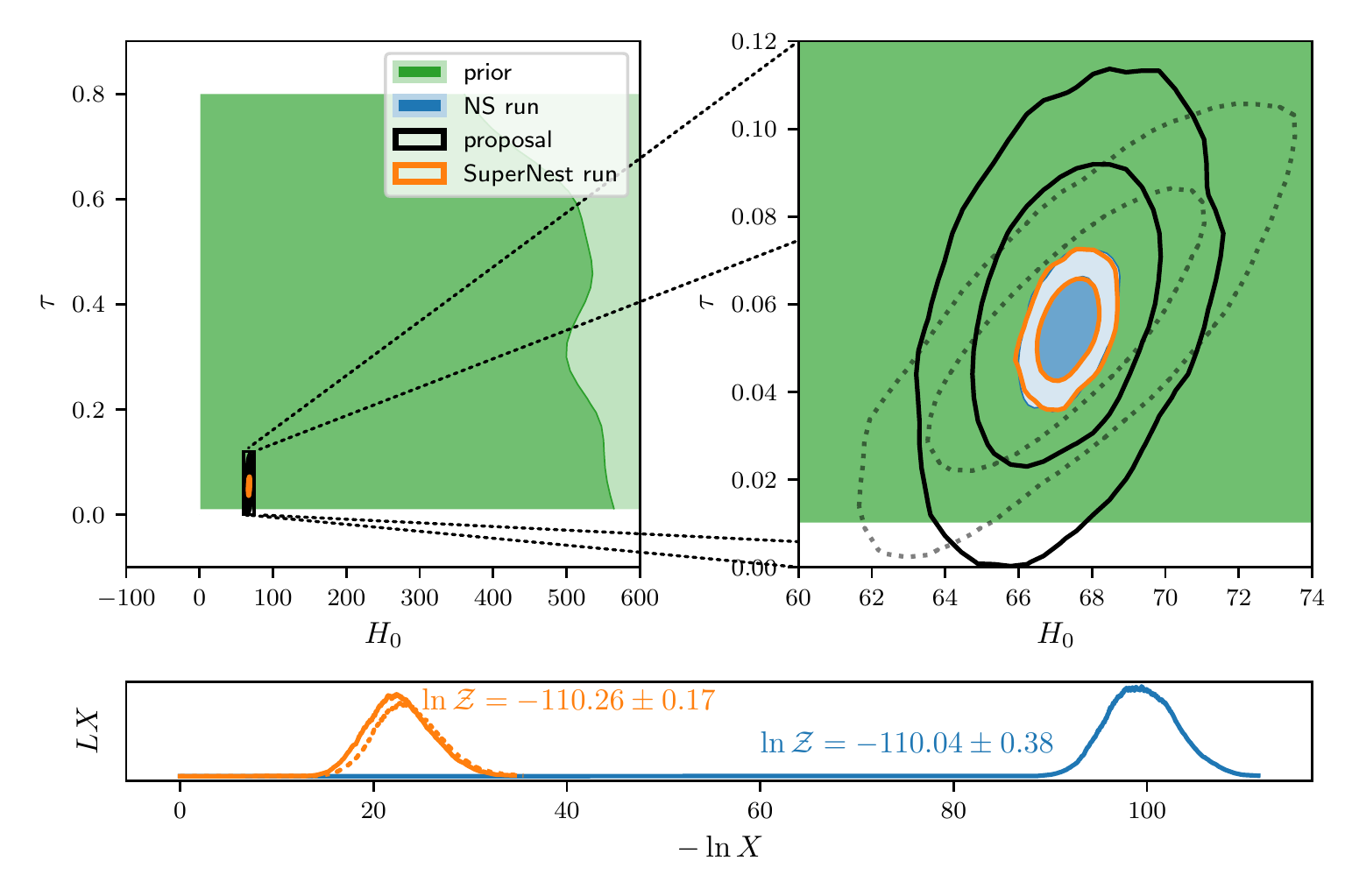}
  \caption{\texttt{supernest} in action. The top panels show the
    parameter spaces of the Hubble constant and optical depth to
    reionisation \((H_0,\tau)\)~\cite{Planck} with all other
    \(\Lambda\)CDM and nuisance parameters marginalised
    out. Contours show the 66\% and 95\% credibility regions of the
    prior (green), and nested sampling posterior (blue), alongside
    the proposal distribution (black) provided to \texttt{supernest}
    which recovers the same posterior (orange) as the original NS
    run. The left hand plot visualised the prior, whilst the right
    hand plot zooms in on the posterior for clarity. The lower panel
    shows the Higson plot~\cite{Higson2018DynamicNS} of the two
    runs. Here the evidences recovered are consistent, and the
    \texttt{supernest} run in orange has a more accurate evidence
    inference, associated with the fact that the compression in
    prior volume \(\ln X\) is substantially lower, meaning the
    nested sampling run terminates in approximately a third the
    time. An alternative proposal distribution with dashed lines is
    also plotted, which recovers an evidence of
    \(\ln\mathcal{Z}=-110.36\pm0.18\), consistent to within sampling
    error.\label{fig:proposal}}
  \end{figure}

  For the purposes of demonstration, we choose a likelihood
  \(\mathcal{L}\) which is Gaussian in the parameters, with parameter
  means and parameter covariances as given by the Planck legacy
  archive chains~\cite{Planck}, and prior given by the default prior
  widths as specified in those chains. This has six cosmological
  parameters and 21 nuisance parameters associated with galactic
  foregrounds and satellite calibration, giving 27 parameters to
  constrain. In its default mode, this therefore gives
  \({n_\mathrm{live}=675}\) and \({n_\mathrm{repeats}=135}\), and for
  this example an \({n_\mathrm{dead}\sim72000}\). This provides a
  proxy for a nested sampling problem of interests to cosmologists,
  whilst being fast enough numerically to explore a variety of
  configurations.

  We run \texttt{PolyChord} wrapped with supernest, both with their
  default settings. \texttt{PolyChord} in addition is set with a
  precision criterion of \(10\%\), i.e.~stopping when the live points
  contain 10\% of the remaining evidence. As a representative example,
  we first choose a Gaussian proposal with the same covariance as the
  parameters, inflated by a factor of \(3^d\) in volume, and offset by
  a random vector drawn from the posterior. This gives a proposal
  which is conservative (much wider than the posterior), and
  approximately correctly centered.

  The performance of this example is shown in \cref{fig:proposal}. We
  can see that the algorithm terminates approximately \(3\) times
  faster, with an error bar that is approximately \(2\) times smaller
  despite the perturbation. Usually to reduce the error bar by a
  factor of \(2\), one would need to increase the number of live
  points by a factor of \(4\). Thus, when precision-normalised
  \texttt{supernest} achieved an order-of-magnitude \(12\times\)
  improvement over the traditional nested sampling run. Further
  improvements to this speed are achievable if a more compact (and
  possibly larger number of) proposal distributions are used.

  As a second example where the covariance of the proposal does not
  align with the covariance of the posterior, we draw an inverse
  Wishart distributed covariance matrix, with $d=27$ degrees of
  freedom and scale matrix equal to the covariance, once again
  inflating the result in volume by $3^d$, and offsetting the result
  by a random vector drawn from the posterior. The performance is
  unaffected to within sampling fluctuation, with the recovered
  evidence equal to within error.

  \section{Conclusions}\label{sec:conclusions}
  We presented a method for improving the performance and accuracy of
  nested sampling using the recent innovations in posterior
  repartitioning of \citet{chen-ferroz-hobson}. The approach functions
  by reducing the effective prior-to-posterior compression that nested
  sampling ``sees'', and hence reducing both the runtime and
  accumulated Poisson error in the evidence estimate. We discussed a
  variety of alternative posterior repartitioning strategies before
  settling on a stochastic superpositional scheme, and demonstrated
  that on a cosmological example even conservative proposals have the
  capacity to speed up nested sampling runs by an order of
  magnitude.

  This performance improvement is achieved by trimming off the long
  prior-to-posterior compression phase seen in the blue but not
  orange curve of the lower panel of \cref{fig:proposal}. The
  remaining majority of the time is spent in traversing the
  typical set. Accelerating the pace at which nested sampling
  does this is a matter of ongoing research.

  This work will be followed by a re-examination of the repartitioning,
  in particular by considering more general cases than the ones
  considered here. In a follow up study we shall provide a more
  mathematical look into how further performance improvements can be
  achieved via modifications to nested samplers, hysteretic priors on
  the choice parameter, as well as exploring the connection between
  prior spaces and Hilbert spaces. Finally, we shall consider a
  reformulation of Bayesian methods using functionals, and explore the
  implications of \emph{consistent partitioning} (a generalisation of
  posterior repartitioning) applied to Bayesian inference in functional
  space.

  \reftitle{References}

  \bibliography{supernest}

\begin{thebibliography}{-------}
\providecommand{\natexlab}[1]{#1}

\bibitem[Chen \em{et~al.}(2019)Chen, Feroz, and Hobson]{chen-ferroz-hobson}
Chen, X.; Feroz, F.; Hobson, M.
\newblock Bayesian automated posterior repartitioning for nested sampling,
  2019,  \href{http://xxx.lanl.gov/abs/arXiv:1908.04655}{{\normalfont
  [arXiv:1908.04655]}}.

\bibitem[Skilling(2006)]{Skilling2006}
Skilling, J.
\newblock Nested sampling for general Bayesian computation.
\newblock {\em Bayesian Anal.} {\bf 2006}, {\em 1},~833--859.
\newblock
  doi:{\changeurlcolor{black}\href{https://doi.org/10.1214/06-BA127}{\detokenize{10.1214/06-BA127}}}.

\bibitem[MacKay(2003)]{importanceOfZ}
MacKay, D.J.C., Information Theory, Inference, and Learning Algorithms;
  Cambridge University press,  2003; chapter 29. Monte-Carlo methods, pp.
  379--380.

\bibitem[Ashton \em{et~al.}(2022)Ashton, Bernstein, Buchner, Chen, Cs{\'a}nyi,
  Fowlie, Feroz, Griffiths, Handley, Habeck, Higson, Hobson, Lasenby,
  Parkinson, P{\'a}rtay, Pitkin, Schneider, Speagle, South, Veitch, Wacker,
  Wales, and Yallup]{NRP}
Ashton, G.; Bernstein, N.; Buchner, J.; Chen, X.; Cs{\'a}nyi, G.; Fowlie, A.;
  Feroz, F.; Griffiths, M.; Handley, W.; Habeck, M.; Higson, E.; Hobson, M.;
  Lasenby, A.; Parkinson, D.; P{\'a}rtay, L.B.; Pitkin, M.; Schneider, D.;
  Speagle, J.S.; South, L.; Veitch, J.; Wacker, P.; Wales, D.J.; Yallup, D.
\newblock Author Correction: Nested sampling for physical scientists.
\newblock {\em Nature Reviews Methods Primers} {\bf 2022}, {\em 2},~44.
\newblock
  doi:{\changeurlcolor{black}\href{https://doi.org/10.1038/s43586-022-00138-2}{\detokenize{10.1038/s43586-022-00138-2}}}.

\bibitem[Arminger and Muth{\'e}n(1998)]{Metropolis}
Arminger, G.; Muth{\'e}n, B.O.
\newblock {A Bayesian approach to nonlinear latent variable models using the
  Gibbs sampler and the metropolis-hastings algorithm}.
\newblock {\em Psychometrika} {\bf 1998}, {\em 63},~271--300.
\newblock
  doi:{\changeurlcolor{black}\href{https://doi.org/10.1007/BF02294856}{\detokenize{10.1007/BF02294856}}}.

\bibitem[Kullback and Leibler(1951)]{Kullback_1951}
Kullback, S.; Leibler, R.A.
\newblock On Information and Sufficiency.
\newblock {\em The Annals of Mathematical Statistics} {\bf 1951}, {\em
  22},~79--86.
\newblock
  doi:{\changeurlcolor{black}\href{https://doi.org/10.1214/aoms/1177729694}{\detokenize{10.1214/aoms/1177729694}}}.

\bibitem[Feroz \em{et~al.}(2009)Feroz, Hobson, and
  Bridges]{Feroz2009MultiNestAE}
Feroz, F.; Hobson, M.P.; Bridges, M.
\newblock MultiNest: an efficient and robust Bayesian inference tool for
  cosmology and particle physics.
\newblock  Monthly notices of the Royal Astronomical Society,  2009.

\bibitem[Handley \em{et~al.}(2015)Handley, Hobson, and Lasenby]{polychord}
Handley, W.J.; Hobson, M.P.; Lasenby, A.N.
\newblock {polychord: next-generation nested sampling}.
\newblock {\em Monthly Notices of the Royal Astronomical Society} {\bf 2015},
  {\em 453},~4384--4398.
\newblock
  doi:{\changeurlcolor{black}\href{https://doi.org/10.1093/mnras/stv1911}{\detokenize{10.1093/mnras/stv1911}}}.

\bibitem[Anstey \em{et~al.}(2021)Anstey, de~Lera~Acedo, and Handley]{hacky}
Anstey, D.; de~Lera~Acedo, E.; Handley, W.
\newblock A general Bayesian framework for foreground modelling and
  chromaticity correction for global 21 cm experiments.
\newblock {\em Monthly Notices of the Royal Astronomical Society} {\bf 2021},
  {\em 506},~2041–2058.
\newblock
  doi:{\changeurlcolor{black}\href{https://doi.org/10.1093/mnras/stab1765}{\detokenize{10.1093/mnras/stab1765}}}.

\bibitem[{Torrado} and {Lewis}(2021)]{2021JCAP...05..057T}
{Torrado}, J.; {Lewis}, A.
\newblock {Cobaya: code for Bayesian analysis of hierarchical physical models}.
\newblock {\em \jcap} {\bf 2021}, {\em 2021},~057,
  \href{http://xxx.lanl.gov/abs/2005.05290}{{\normalfont
  [arXiv:astro-ph.IM/2005.05290]}}.
\newblock
  doi:{\changeurlcolor{black}\href{https://doi.org/10.1088/1475-7516/2021/05/057}{\detokenize{10.1088/1475-7516/2021/05/057}}}.

\bibitem[{Lewis} and {Bridle}(2002)]{2002PhRvD..66j3511L}
{Lewis}, A.; {Bridle}, S.
\newblock {Cosmological parameters from CMB and other data: A Monte Carlo
  approach}.
\newblock {\em \prd} {\bf 2002}, {\em 66},~103511,
  \href{http://xxx.lanl.gov/abs/astro-ph/0205436}{{\normalfont
  [arXiv:astro-ph/astro-ph/0205436]}}.
\newblock
  doi:{\changeurlcolor{black}\href{https://doi.org/10.1103/PhysRevD.66.103511}{\detokenize{10.1103/PhysRevD.66.103511}}}.

\bibitem[{Lewis}(2013)]{2013PhRvD..87j3529L}
{Lewis}, A.
\newblock {Efficient sampling of fast and slow cosmological parameters}.
\newblock {\em \prd} {\bf 2013}, {\em 87},~103529,
  \href{http://xxx.lanl.gov/abs/1304.4473}{{\normalfont
  [arXiv:astro-ph.CO/1304.4473]}}.
\newblock
  doi:{\changeurlcolor{black}\href{https://doi.org/10.1103/PhysRevD.87.103529}{\detokenize{10.1103/PhysRevD.87.103529}}}.

\bibitem[{Brinckmann} and {Lesgourgues}(2019)]{2019PDU....24..260B}
{Brinckmann}, T.; {Lesgourgues}, J.
\newblock {MontePython 3: Boosted MCMC sampler and other features}.
\newblock {\em Physics of the Dark Universe} {\bf 2019}, {\em 24},~100260,
  \href{http://xxx.lanl.gov/abs/1804.07261}{{\normalfont
  [arXiv:astro-ph.CO/1804.07261]}}.
\newblock
  doi:{\changeurlcolor{black}\href{https://doi.org/10.1016/j.dark.2018.100260}{\detokenize{10.1016/j.dark.2018.100260}}}.

\bibitem[{Zuntz} \em{et~al.}(2015){Zuntz}, {Paterno}, {Jennings}, {Rudd},
  {Manzotti}, {Dodelson}, {Bridle}, {Sehrish}, and
  {Kowalkowski}]{2015A&C....12...45Z}
{Zuntz}, J.; {Paterno}, M.; {Jennings}, E.; {Rudd}, D.; {Manzotti}, A.;
  {Dodelson}, S.; {Bridle}, S.; {Sehrish}, S.; {Kowalkowski}, J.
\newblock {CosmoSIS: Modular cosmological parameter estimation}.
\newblock {\em Astronomy and Computing} {\bf 2015}, {\em 12},~45--59,
  \href{http://xxx.lanl.gov/abs/1409.3409}{{\normalfont
  [arXiv:astro-ph.CO/1409.3409]}}.
\newblock
  doi:{\changeurlcolor{black}\href{https://doi.org/10.1016/j.ascom.2015.05.005}{\detokenize{10.1016/j.ascom.2015.05.005}}}.

\bibitem[{Planck Collaboration}(2020)]{Planck}
{Planck Collaboration}.
\newblock {Planck 2018 results. VI. Cosmological parameters}.
\newblock {\em \aap} {\bf 2020}, {\em 641},~A6,
  \href{http://xxx.lanl.gov/abs/1807.06209}{{\normalfont
  [arXiv:astro-ph.CO/1807.06209]}}.
\newblock
  doi:{\changeurlcolor{black}\href{https://doi.org/10.1051/0004-6361/201833910}{\detokenize{10.1051/0004-6361/201833910}}}.

\bibitem[{Edward Higson and Will Handley and Mike Hobson and Anthony N.
  Lasenby}(2018)]{Higson2018DynamicNS}
{Edward Higson and Will Handley and Mike Hobson and Anthony N. Lasenby}.
\newblock {Dynamic nested sampling: an improved algorithm for parameter
  estimation and evidence calculation}.
\newblock {\em Statistics and Computing} {\bf 2018}, pp. 1--23.

\end{thebibliography}

\end{document}